\RequirePackage[l2tabu, orthodox]{nag}

\documentclass{webofc}
\usepackage[varg]{txfonts}   
\usepackage[utf8]{inputenc}
\usepackage[T1]{fontenc}
\usepackage[english]{babel}
\usepackage{amsmath}
\usepackage{amssymb}
\usepackage{braket}
\usepackage{bm}
\usepackage{hyperref}
\usepackage{latexsym}
\usepackage{graphicx,caption}
\usepackage{rotating}
\usepackage{grffile}
\usepackage{color}
\usepackage{braket}
\usepackage{amssymb}
\usepackage{listings}
\usepackage{placeins}
\usepackage{multirow}
\usepackage[normalem]{ulem}

\newcommand{\FFT}{\mathop{\text{FFT}}}
\newcommand{\IFFT}{\mathop{\text{IFFT}}}
\renewcommand{\i}{\mathop{\text{i}}}
\newcommand{\ig}[1]{\includegraphics[width=.49\textwidth]{#1}}

\usepackage{mdwlist}
{
\begin{basedescript}{
\desclabelstyle{\nextlinelabel}
\desclabelwidth{2em}}}
{
\end{basedescript}
}

\newcounter{bla}

\usepackage{todonotes}
\setuptodonotes{inline}

\begin{document}

\title{QuantumFDTD - A computational framework for the relativistic Schrödinger equation}

\author{
    \firstname{Rafael} \lastname{L.~Delgado} \inst{1,2,3} \fnsep\thanks{\email{rafael.delgado@upm.es}}
    \and
    \firstname{Sebastian} \lastname{Steinbei{\ss}er} \inst{3,4} \fnsep\thanks{\email{sebastian.steinbeisser@tum.de}}
    \and
    \firstname{Michael} \lastname{Strickland} \inst{5} \fnsep\thanks{\email{mstrick6@kent.edu}}
    \and
    \firstname{Johannes~H.} \lastname{Weber} \inst{6} \fnsep\thanks{\email{johannes.weber@physik.hu-berlin.de}}
}

\institute{
    ETSIS de Telecomunicación (UPM), Campus Sur, C/Nikola Tesla, s/n 28031 Madrid, Spain
    \and
    INFN-Firenze, Via Giovanni Sansone, 1, 50019 Sesto Fiorentino FI, Italy (moved to UPM)
    \and
    Physik Department, Technische Universität München, James-Franck-Straße~1, \\
    D-85748 Garching b.\ München, Germany
    \and
    Leibniz-Rechenzentrum der Bayerischen Akademie der Wissenschaften, Boltzmannstraße~1, \\
    D-85748 Garching b.\ München, Germany
    \and
    Department of Physics, Kent State University, Kent, OH 44242 USA
    \and
    Institut für Physik \& IRIS Adlershof, Humboldt-Universität zu Berlin, Zum Großen Windkanal~6, \\
    D-12489 Berlin, Germany
}

\abstract{
We extend the publicly available quantumfdtd code.
It was originally intended for solving the time-independent three-dimensional Schrödinger equation via the finite-difference time-domain (FDTD) method and for extracting the ground, first, and second excited states.
We (a) include the case of the relativistic Schrödinger equation and (b) add two optimized FFT-based kinetic energy terms for the non-relativistic case.
All the three new kinetic terms are computed using Fast Fourier Transform (FFT).
We release the resulting code as version 3 of quantumfdtd.
Finally, the code now supports arbitrary external file-based potentials and the option to project out distinct parity eigenstates from the solutions.
Our goal is quark models used for phenomenological descriptions of QCD bound states, described by the three-dimensional Schrödinger equation.
However, we target any field where solving either the non-relativistic or the relativistic three-dimensional Schrödinger equation is required.
}

\maketitle

\section{Introduction}
In our work~\cite{Delgado:2020ozh} we extended the \verb+quantumfdtd+ software package~\cite{Strickland:2009ft, Dumitru:2009ni, Dumitru:2009fy, Margotta:2011ta}.
Besides containing a detailed study about numerical stability of the new \verb+quantumfdtd+ code, Ref.~\cite{Delgado:2020ozh} is a reference manual for the code.

The legacy code is a parallelized solver for the three-dimensional non-relativistic Schrödinger equation with a selection of analytical real and complex potentials.
The solver is based on the finite-difference time-domain (FDTD) method.
Via an iterative method (inherited from the legacy \verb+quantumfdtd+ code~\cite{Strickland:2009ft, Dumitru:2009ni, Dumitru:2009fy, Margotta:2011ta}), the code is able to extract the ground, first, and second excited states.
The excited states are determined from multiple nearby temporal snapshots of the iterative procedure that are consecutively subjected to orthogonal projection.
The analytical hard-coded potentials could be easily extended by writing the appropriate code inside the \verb+potentials.cpp+ source file.
The chosen FDTD method implies that the boundary conditions in the edge of the 3D lattice cube are Dirichlet ones, with the wavefunction, $\Psi$, vanishing on the boundary, $\Psi(\text{boundary}) = 0$.
The only library that is required to run the code is \verb+MPI+, for the parallelization.

In our extension of the \verb+quantumfdtd+ package three new kinetic terms are included.
These are based on the Fast Fourier Transform (FFT), that requires periodic boundary conditions in the edge of the 3D lattice cube.
For the non-relativistic and FFT-based solvers, most of the differences from the FDTD results come from this difference in the boundary conditions, as was extensively discussed on our work~\cite{Delgado:2020ozh}.
Part of such a study will be also reviewed here.

We consider two non-relativistic FFT-based kinetic terms, the difference between them being the usage of the Symanzik effective field theory~\cite{Symanzik:1983dc, Symanzik:1983gh} for reducing the discretization errors.
The third FFT-based kinetic term is the relativistic one, intended for solving the 3D relativistic Schrödinger equation.
For the computation of the FFT we require our code to be linked to the \verb+FFTW_MPI+ version~3 library.\footnote{%
\href{http://fftw.org/}{http://fftw.org/} and Ref.~\cite{fftw3}.}

The target of this software is quark models used for phenomenological description of QCD bound states.
Most of these models are described by the three-dimensional Schrödinger equation with different potentials.
In particular, these models have successfully described below-threshold charmonium production and bottomonium spectra and have helped to establish confidence in QCD as the first-principles description of hadronic matter~\cite{Ebert:2002pp, Eichten:2007qx, Segovia:2016xqb}.
Non-relativistic effective field theory methods have been used for a first-principles approach to potential-based non-relativistic QCD (pNRQCD)~\cite{Brambilla:1999xf, Brambilla:1999qa}.
In order to describe quarkonium evolution in the quark-gluon plasma, these potential models have been extended to finite temperature~\cite{Laine:2006ns, Brambilla:2008cx, Brambilla:2010xn} and non-equilibrium~\cite{Dumitru:2007hy, Burnier:2009yu, Dumitru:2009fy, Dumitru:2009ni, Guo:2018vwy}.
In the latter case, the potentials are no longer real-valued or spherically symmetric.
In the full non-equilibrium case, a full three-dimensional solver, like the one described in this work, is necessary.

Finally, for integration into a lattice field theory workflow (and for the general convenience of the user independently of the research field) we have extended the code by allowing for loading arbitrary complex potentials from ASCII files with numerical values.
We now require the GNU Scientific Library (\verb+GSL+),\footnote{%
\href{https://www.gnu.org/software/gsl/}{https://www.gnu.org/software/gsl/} and Ref.~\cite{gsl}} with the CBLAS link.\footnote{%
\href{https://www.netlib.org/blas/}{https://www.netlib.org/blas/} and Refs.~\cite{blas1,blas2,blas3}}
We also include several post-processing scripts.
The most powerful ones aim to aid the extraction of excited states by exploiting their symmetries.
For running these scripts, Python~3 is required.\footnote{%
\href{https://www.python.org/downloads/}{https://www.python.org/downloads/}}

\section{New kinetic terms and iterative procedure}
The Schrödinger Hamiltonian
\begin{equation}
    \label{eq:eq_schro}
    H = H_{\text{K}} + V(\vec{r}) \,,
\end{equation}
can be split into a kinetic piece $H_{\text{K}}$ and a potential $V(\vec{r})$.
The non-relativistic kinetic term $H_{\text{K}}^{\text{nr}}$, already implemented (in its FDTD discretized version) in the legacy \verb+quantumfdtd+ code, and the relativistic one $H_{\text{K}}^{\text{rel}}$ are given, respectively, by
\begin{equation}
    \label{eq:HKnr}
    H_{\text{K}}^{\text{nr}} = \sum\limits_{i=1,2,3} \frac{p_{i}^{2}}{2m}, \quad
    H_{\text{K}}^{\text{rel}} = \sqrt{ m^{2} + \sum\limits_{i=1,2,3} p_{i}^{2} } \,,
\end{equation}
where $\vec{p} = (p_{1},p_{2},p_{3})$ is the spatial three momentum.

We provide the following four implemented kinetic terms:
the original one $H_{\text{K}}^{(0)}$, based on the FDTD discretization,
\begin{equation}
    \label{eq:HK0}
    H_{\text{K}}^{(0)} \Psi = - \frac{1}{2M} \sum\limits_{l = \pm 1}^{\pm 3} \frac{1}{2A} \frac{\Psi(\vec{r} + \hat{e}_{l}) - \Psi(\vec{r})}{A} \,,
\end{equation}
and three new kinetic terms, $H_{\text{K}}^{(1)}$, $H_{\text{K}}^{(2)}$ (both non-relativistic) and $H_{\text{K}}^{(3)}$ (relativistic),
\begin{align}
    & H_{\text{K}}^{(1)} \Psi = \frac{1}{2A^{2}M N^{3}} \cdot \IFFT\left[ \sum\limits_{l=1,2,3} \left(k_{l}\right)^{2} \cdot \FFT[\Psi] \right] \,, \label{eq:HK1} \\
    & H_{\text{K}}^{(2)} \Psi = \frac{1}{2A^{2}M N^{3}} \cdot \IFFT\left[ 4\sum\limits_{l=1,2,3} \sin^{2}\!\left(\frac{k_{l}}{2}\right) \cdot \FFT[\Psi] \right] \,, \label{eq:HK2} \\
    & H_{\text{K}}^{(3)} \Psi = \frac{1}{A N^{3}} \cdot \IFFT\left[ \sqrt{4\sum\limits_{l=1,2,3} \sin^{2}\!\left(\frac{k_{l}}{2}\right) + (AM)^{2}} \cdot \FFT[\Psi] \right] \,.\label{eq:HK3}
\end{align}
Here, $\hat{e}_{l}$ denotes the (signed) vector in the $l$ direction ($l = \pm 1,\pm 2, \pm 3$) of one unit of the lattice spacing $A$ and $k_{l} \equiv A p_{l} \in \{ -\pi + 2\pi/N, \dots, 0, 2\pi/N, 4\pi/N, \dots, \pi \}$ are the lattice momentum coordinates.
The $\FFT$ and $\IFFT$ operators denote the Fast Fourier Transform and Inverse Fast Fourier Transform, that are computed via the \verb+FFTW_MPI+ version~3 library~\cite{fftw3}, as stated in the previous section.
Finally, $\Psi(\vec{r}) \equiv \Psi[(Ax_{1}, Ax_{2}, Ax_{3})]$ for convenience reasons.

For solving the Schrödinger equation, an iterative method inherited from the legacy \verb+quantumfdtd+ code~\cite{Strickland:2009ft, Dumitru:2009ni, Dumitru:2009fy, Margotta:2011ta} is used.
Since this method has not been modified from the legacy code, the reader is referred to~\cite{Strickland:2009ft} for its description.
In brief, it is based on a Wick rotation of the temporal variable $\i t \to \tau$, so that the Schrödinger equation turns into
\begin{equation}\label{eq_iterat}
    -\frac{\partial}{\partial\tau} \Psi(\vec{r},\tau) = H_{\text{K}} \Psi(\vec{r},\tau) \,.
\end{equation}
Its general solution
\begin{equation}
    \Psi(\vec{r},\tau) = \sum\limits_{n=0}^{\infty} a_{n} \Psi_{n}(\vec{r}) e^{-E_{n}\tau} \,,
\end{equation}
allows for the extraction of both the wavefunctions $\Psi_{n}$ and energies $E_{n}$ of the ground $n=0$ and excited $n \geq 1$ states, once the intermediate wavefunctions $\Psi(\vec{r},\tau)$ have been obtained.
In order to extract such a solution $\Psi(\vec{r},\tau)$, we discretize the temporal derivative in Eq.~\ref{eq_iterat}.
For details see Ref.~\cite{Delgado:2020ozh}.

\section{Results}

\begin{figure}[ht]
\centering
\ig{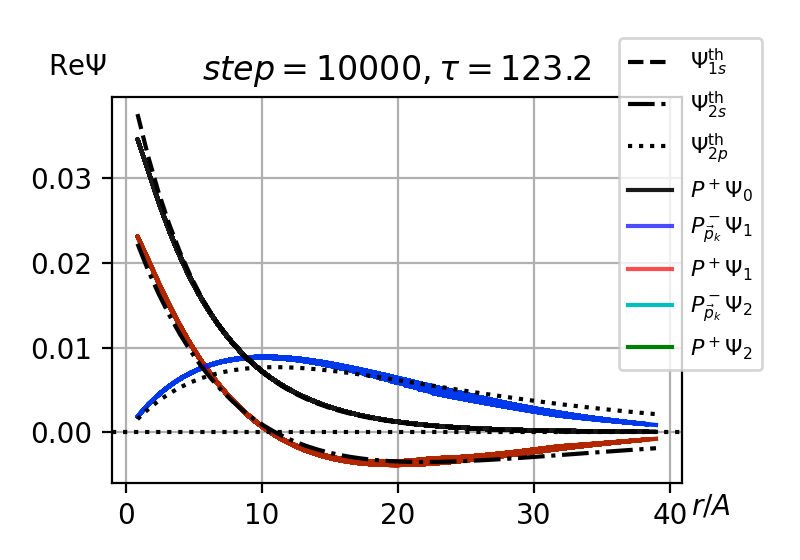}
\hfill%
\ig{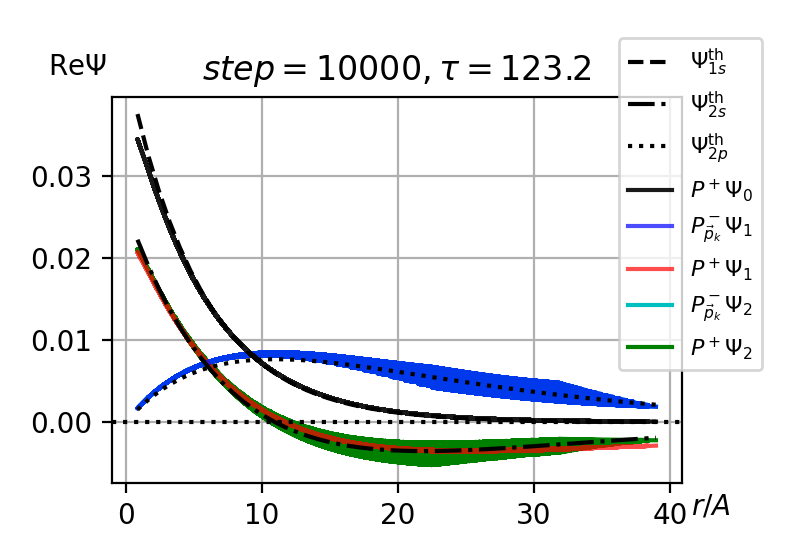}
\caption{\label{fig:finite_volume_effects}%
Figures from Ref.~\cite{Delgado:2020ozh}: Coulomb potential, $N=64$, $A=0.12$~fm, $M=0.3$~GeV.
From left to right: wavefunctions of $H_{\text{K}}=H_{\text{K}}^{(0)}$ and $H_{\text{K}}^{(2)}$ at $\tau=123.2$ (non-relativistic kinetic terms).
Observe the different finite volume effects between the FDTD (left) and FFT (right) solvers.
The behavior of the $H_{\text{K}}^{(1)}$ solver is almost indistinguishable from the $H_{\text{K}}^{(2)}$ one (see Ref.~\cite{Delgado:2020ozh}).}
\end{figure}

Ref.~\cite{Delgado:2020ozh} includes a comprehensive study of all the relativistic and non-relativistic kinetic terms, $H_{\text{K}}^{(i)}$ ($i=0,\dots,3$).
Thus, we summarize here only some of the most important results.

For the non-relativistic kinetic terms, we use Coulomb potential (for which the analytical solution is known), a lattice spacing of $A=0.12$~fm, and a reduced mass of $M=0.3$~GeV.
Our goal is demonstrate the capabilities of the code by reproducing the analytically known wavefunctions and eigenenergies.
Of course, both discretization errors and finite volume effects will lead to deviations.
Finite volume effects, in particular, are expected to produce a discrepancy between $H_{\text{K}}^{(0}$ (based on the FDTD method) and $H_{\text{K}}^{(1)}$, $H_{\text{K}}^{(2)}$ (based on FFT), due to the different boundary conditions: Dirichlet ones for $H_{\text{K}}^{(0)}$ and periodic ones for $H_{\text{K}}^{(1)}$ and $H_{\text{K}}^{(2)}$.

\begin{figure}[ht]
\centering
\ig{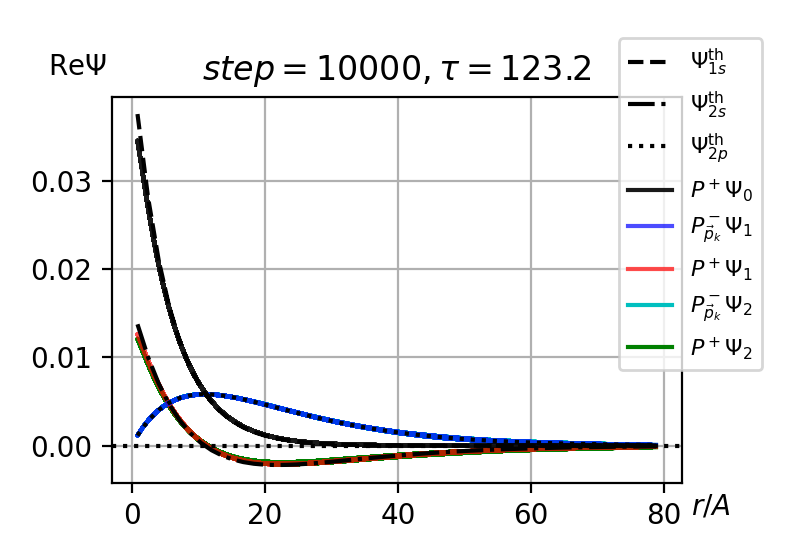}
\hfill%
\ig{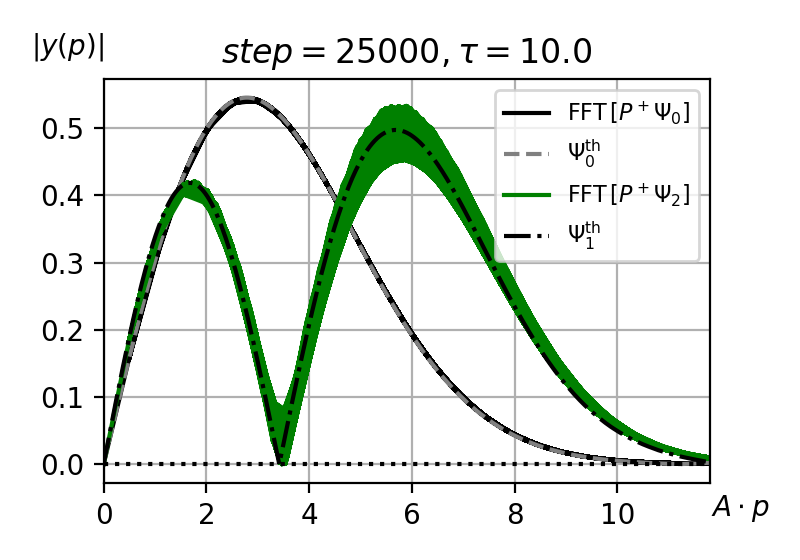}
\caption{\label{fig:correct_behaviour_high_N}%
Figures from Ref.~\cite{Delgado:2020ozh}.
Left: as in Fig.~\ref{fig:finite_volume_effects} but with $N=128$.
The wavefunction from the kinetic terms $H_{\text{K}}^{(0)}$ and $H_{\text{K}}^{(1)}$ are almost indistinguishable~\cite{Delgado:2020ozh}.\\
Right: relativistic kinetic term $H_{\text{K}}^{(3)}$ with the modified harmonic oscillator potential (Ref.~\cite{Li:2005bd}) momentum space wavefunctions with the relativistic kinetic term $H_{\text{K}}^{(3)}$.
Here, $y(p) = \sqrt{4\pi} p \Psi(p)$ in order to compare to the theoretical predictions following~\cite{Li:2005bd}.}
\end{figure}

In Fig.~\ref{fig:finite_volume_effects} we can see the analytical ground, first, and second excited states $\Psi_{1s}^{\text{th}}$, $\Psi_{2s}^{\text{th}}$, $\Psi_{2p}^{\text{th}}$ of the Coulomb potential.
In the same plots, we show parity projections
\begin{align}
&P^{\pm} \Psi(\vec{r}) = \frac{1}{2}\Big(\Psi(\vec{r})\pm\Psi(-\vec{r})\Big) \,, \\
&\begin{aligned}
P^{-}_{\vec{p}_{k}}(x_1,x_2,x_3) = \frac{1}{2}\Big(&\Psi(x_1,x_2,x_3) + \Psi(-x_1,-x_2,x_3) \\
&- \Psi(x_1,x_2,-x_3) - \Psi(-x_1,-x_2,-x_3)\Big) \,,
\end{aligned}
\end{align}
of the wavefunctions computed by \verb+quantumfdtd+ with the $H_{\text{K}}^{(0)}$ kinetic term (FDTD based, left) and $H_{\text{K}}^{(2)}$ (FFT based, right).
Note that, because of the structure of the exact (analytical) solution, $P^{+} \Psi_{ns}(\vec{r}) = \Psi_{ns}(\vec{r})$, $P^{-}_{\vec{p}_{k}} \Psi_{ns}(\vec{r}) = 0$, $P^{+} \Psi_{n'p}(\vec{r}) = 0$ and $P_{\vec{p}_{k}} \Psi_{n'p}(\vec{r}) = \Psi_{n'p}(\vec{r})$, where $n=1,2,3,\dots$ and $n'=2,3,\dots$.
These parity projectors, that have been implemented~\cite{Delgado:2020ozh} in the Python~3 scripts included with \verb+quantumfdtd+, allow for a more precise extraction of the ground, first, and second excited states.
Indeed, the results shown in Fig.~\ref{fig:finite_volume_effects} are very close to the analytically known solution.

In the left panel of Fig.~\ref{fig:finite_volume_effects} (FDTD based, Dirichlet boundary conditions) the wavefunctions decay too fast near the edge (expected with Dirichlet boundary conditions).
On the right (periodic boundary conditions) the wavefunctions show artifacts near the boundary.
These artifacts are due to the breaking of spherical symmetry.
This is also expected for periodic boundary conditions with a non-spherically symmetric, cubic, lattice.
Finally, on both plots there are artifacts near the origin $r/A \to 0$, where discretization effects are more prominent (and the Coulomb potential has a pole).
The wavefunction of $H_{\text{K}}^{(1)}$ is almost indistinguishable from that of $H_{\text{K}}^{(2)}$ (see Ref.~\cite{Delgado:2020ozh}).

On the left of Fig.~\ref{fig:correct_behaviour_high_N} we show the non-relativistic wavefunction obtained using a finer lattice $N=128$.
The wavefunctions of $H_{\text{K}}^{(0)}$, $H_{\text{K}}^{(1)}$, and $H_{\text{K}}^{(2)}$ are almost indistinguishable and reproduce almost exactly the theoretical solutions $\Psi_{1s}^{\text{th}}$, $\Psi_{2s}^{\text{th}}$, $\Psi_{2p}^{\text{th}}$.
The only difference is a small artifact very close to the origin $r/A = 0$, where discretization effects (finite $A$) are most prominent.

Finally, as a means to check the relativistic kinetic term $H_{\text{K}}^{(3)}$, we included the right plot of Fig.~\ref{fig:correct_behaviour_high_N} to check a specially modified version of the relativistic harmonic oscillator potential for which an approximate analytical solution exists~\cite{Li:2005bd}.
In momentum space (chosen by Ref.~\cite{Li:2005bd}) our solver is able to reproduce the analytical result of~\cite{Li:2005bd} for the ground, $1s$, and first excited, $2s$, states to sub percent levels.
Furthermore, our code is not restricted to $s$-wave states as is the analytical procedure of Ref.~\cite{Li:2005bd}.
For instance, we can compute the $2p$ state, as shown in~\cite{Delgado:2020ozh}.
Our goal here is a direct comparison between our code~\cite{Delgado:2020ozh} and the analytical procedure~\cite{Li:2005bd} where the latter one is applicable, so that we can cross-check the relativistic kinetic term $H_{\text{K}}^{(3)}$.
As shown in Fig.~\ref{fig:correct_behaviour_high_N}, our code behaves correctly at sub percent level in this application of the relativistic solver.

\section{Conclusions}
In this work we have extended the legacy \verb+quantumfdtd+ code to support the relativistic Schrödinger equation and we have added two new solvers of the non-relativistic one.
The new solvers are based on the Fast Fourier Transform (FFT) and have periodic boundary conditions instead of Dirichlet ones.
Furthermore, we have performed a comprehensive study of the stability of the solvers and compared them against analytical (exact) solutions of the non-relativistic Coulomb potential and a relativistic particle in a harmonic oscillator potential.
Furthermore, we have extended the code to increase its usability by the lattice field theory community and by the larger scientific community interested in solving the 3D relativistic and non-relativistic Schrödinger equation.

There are a number of possible extensions to the present version of \verb+quantumfdtd+ that are beyond the scope of this work.
From easiest to most difficult: variations of the coupling strength of the hard-coded potentials; new projection operators (for instance, $d$-wave states); implementing a suitable binary format for the wavefunctions; an automatized algorithm to look for saddle points of $E_{i}(\tau)$ ($i=0,1,2$), that is, automatizing the detection of the stabilization of the wavefunctions and eigenenergies; an implicit Crank-Nicolson method for the solver (difficult for the relativistic kinetic term, which is non-linear); and permitting the (anti)-symmetric solution of a \emph{two}-particle system in a background potential with an interaction potential.

\section*{Acknowledgments}
R.L.D. was financially supported by the Ram{\'o}n Areces Foundation, the INFN post-doctoral fellowship AAOODGF-2019-0000329 and the Spanish grants MICINN: PID2019-108655GB-I00/AEI/10.13039/501100011033, PID2021-124473NB-I00.
M.S. was supported by the U.S. Department of Energy, Office of Science, Office of Nuclear Physics Award No.~DE-SC0013470.
J.H.W.’s research is funded by the Deutsche Forschungsgemeinschaft (DFG, German Research Foundation)---Projektnummer 417533893/GRK2575 ``Rethinking Quantum Field Theory''.
We acknowledge the support by the Deutsche Forschungsgemeinschaft (DFG, German Research Foundation) under Germany's Excellence Strategy -- EXC-2094 -- 390783311 via the Excellence Cluster "ORIGINS".
The simulations have been carried out on the computing facilities of the Computational Center for Particle and Astrophysics (C2PAP) and the Leibniz Supercomputing Center (SuperMUC), on the local theory cluster (T30 cluster) of the Physics Department of the Technische Universität München (TUM), and on local computing facilities at the INFN-Firenze.


\bibliography{references}

\end{document}